\documentclass[namedreferences]{solarphysics}

\usepackage{graphicx}
\usepackage{color}
\usepackage{epstopdf}
\usepackage[optionalrh,solaromanenum]{spr-sola-addons} 
\usepackage[pdfborder={0 0 0 },urlcolor=blue,breaklinks]{hyperref}
\usepackage{url}

\ifx \adsurl    \undefined \def \adsurl #1{\href{http://adsabs.harvard.edu/abs/#1}{\textsf{ADS}}}\fi
\ifx \doiurl    \undefined \def \doiurl #1{\href{http://dx.doi.org/#1}{\textsf{doi}}}\fi
\ifx \arxivurl  \undefined \def \arxivurl #1{\href{http://arxiv.org/abs/#1}{\textsf{arXiv}}}\fi
\ifx \refurl  \undefined \def \refurl #1{\href{#1}{\textsf{url}}}\fi

\newcommand{\TD}[2]{{\frac{\mathrm{d}{#1}} {\mathrm{d}{#2}}}}

\newcommand{\vet}[1]{{\mbox{\boldmath${\mathrm{#1}}$} }}  

\newcommand{\rmG}{{{\mathrm{G}}}}

\newcommand{\br}{\vet r}
\newcommand{\bxi}{\vet \xi}

\newcommand{\bA}{\vet A}

\newcommand{\bS}{\vet S}

\newcommand{\rmd}{{{\mathrm{d}}}}
\newcommand{\rmR}{{{\mathrm{R}}}}
\newcommand{\rmi}{{{\mathrm{i}}}}

\newcommand{\eg}{{\textit{e.g.}}}
\newcommand{\ie}{{\textit{i.e.}}}

\date{{Received: 3 october2013 / Accepted: 2 December 2013 / Published online:  12 December 2013}}

\begin{document}
\begin{opening}
    \title{Propagating Linear Waves in Convectively Unstable Stellar Models: a Perturbative Approach.}
    \author{E.~\surname{Papini}$^{1}$\sep
            L.~\surname{Gizon}$^{1,2}$\sep
            A.C.~\surname{Birch}$^{1}$      
           }
    \institute{$^{1}$
            Max-Planck-Institut f\"ur Sonnensystemforschung, Justus-von-Liebig-Weg 3, 37077 G\"ottingen, Germany \\e-mail: \href{mailto:papini@mps.mpg.de}{papini@mps.mpg.de} 
            \\e-mail: \href{mailto:birch@mps.mpg.de}{birch@mps.mpg.de}
            \\
            $^{2}$Institut f\"ur Astrophysik, Georg-August-Universit\"at G\"ottingen, 37077 G\"ottingen, Germany 
            \\e-mail: \href{mailto:gizon@astro.physik.uni-goettingen.de}{gizon@astro.physik.uni-goettingen.de}
              }

    \runningauthor{E. Papini {\it et al.}}
    \runningtitle{Linear Waves in Convectively Unstable Stellar Models: a Perturbative Approach.}

\begin{abstract}
Linear time-domain simulations of acoustic oscillations are unstable in the stellar convection zone. To overcome this problem it is customary to compute the oscillations of  a stabilized background stellar model. The stabilization, however, affects the result. Here we propose to use a perturbative approach (running the simulation twice) to approximately recover the acoustic wave field, while preserving seismic reciprocity. To test the method we considered a 1D standard solar model. We found that the mode frequencies of the (unstable) standard solar model are well approximated by the perturbative approach within $1$~$\mu$Hz for low-degree modes with frequencies near $3$~mHz. We also show that the perturbative approach is appropriate for correcting rotational-frequency kernels. Finally, we comment that the method can be generalized to wave propagation in 3D magnetized stellar interiors because the magnetic fields have stabilizing effects on convection.
\end{abstract}
\end{opening}

\section{Time-Domain Simulations of Linear Oscillations}

Helioseismology is used to study complex phenomena in the solar interior and atmosphere, such as flows and magnetic heterogeneities that cover many temporal and spatial scales. Numerical simulations of wave propagation are a crucial 
tool for modeling and interpreting helioseismic observations. The same simulations should find applications in the study of stellar oscillations as well (low-degree modes).

Acoustic waves in the Sun have very low amplitudes compared with those of the background \cite{Christensen-Dalsgaard2002} and thus can be treated as weak perturbations with respect to a background reference model.

The linearized oscillation equations can be solved as an eigenvalue problem (\eg~\opencite{Monteiro2009}; \opencite{Christensen-Dalsgaard2008}) or through time-domain simulations. 
Here we are concerned with the time-domain simulations.
Several linear codes exist in the framework of helioseismology 
(\eg~\opencite{Khomenko2006}; \opencite{Cameron2007}; \opencite{Hanasoge2007}; \opencite{Parchevsky2007}; \opencite{Hartlep2008}). 
Time-domain codes are particularly suited for problems in local helioseismology (see, \eg~\opencite{Gizon2010}; \opencite{Gizon2013}).
They are also useful for the study of wave propagation in slowly evolving backgrounds, \eg~through large-scale convection and magnetic activity.

\subsection{Background Stabilization in Time-Domain Simulations}

A stable background model is required to prevent numerical solutions that grow exponentially with time.
Stellar models, however, always contain dynamical instabilities, which can be of hydrodynamic and/or magnetic nature. These instabilities must be removed.
The main source of instability in stars is convection.
Some magnetic configurations can also be unstable (\eg~\opencite{Tayler1972}), 
although the magnetic field often has a stabilizing effect on convection \cite{Gough1966,Moreno-Insertis1989}.

For the hydrodynamic case, the Schwarzschild criterion \cite{Schwarzschild1906} for local convective stability is
\begin{equation}
  \label{eq:schwarzshild}
  \frac{1}{\Gamma} - \TD{\ln \rho}{\ln P}  < 0,
\end{equation}
where $\rho$, $P$, and $\Gamma$ are density, pressure, and first adiabatic exponent.
This criterion for convective stability can be reformulated to explicitly include gravity [$\vet g(\br)$] by introducing the {Brunt\textendash V\"{a}is\"{a}l\"{a}} or {buoyancy frequency} [$N$]:   
\begin{equation}
\label{eq:SchwarzA}
 N^2 \equiv - \vet g \cdot \vet A > 0,
 \end{equation}
 where 
 \begin{equation}
 \vet A(\br)  \equiv \nabla \ln \rho - \frac{1}{\Gamma} \nabla \ln P
\end{equation} 
is the Schwarzschild discriminant at position $\br$. 
In the solar case, the square of the buoyancy frequency is marginally negative in the convection zone, except for a 
strong negative peak in the highly superadiabatic layer just below the surface.
\begin{figure}
 \hspace*{.7 cm}
 \includegraphics[width=.9\textwidth]{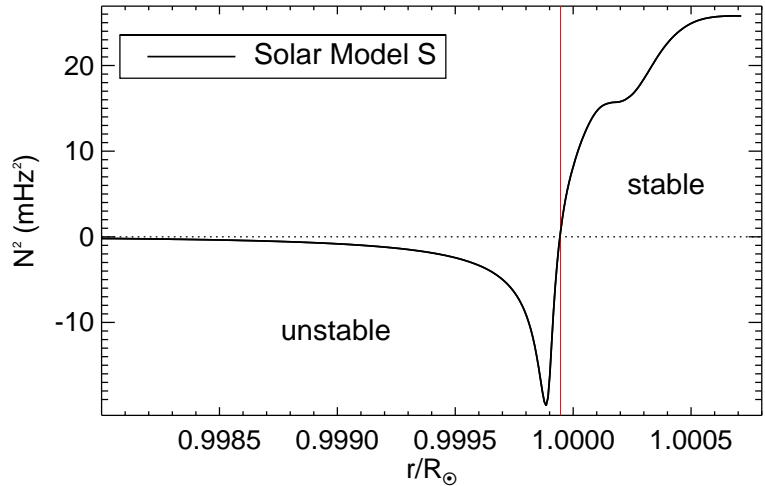}
 \caption[Stabilizing solar Model S]{
    \begin{footnotesize}
    Square of the buoyancy frequency for Model S (solid black line) 
    in the upper part of the convection zone and the atmosphere: the negative peak corresponds to the superadiabatic layer, located just below the photosphere. The vertical red line divides stable and unstable zones.
    \end{footnotesize}}
\label{fig:N2MS}
\end{figure}
Figure \ref{fig:N2MS} shows the squared buoyancy frequency in the upper part of the solar convection zone for Model S \cite{Christensen-Dalsgaard1996}. 

To perform time-domain simulations we need to modify the model in order to obtain a non-negative $N^2$ everywhere. 
Various examples can be found in the literature.
\inlinecite{Hanasoge2006} replaced the near-surface layer above $0.98 \rmR_\odot$ with an empirical model that satisfies convective stability while preserving hydrostatic equilibrium, allowing stable simulations to be extended over a temporal window of several days.
\inlinecite{Hartlep2008} neglected the terms containing $\vet A$ in the momentum equation because they did not affect the frequencies in their range of investigation.
\inlinecite{Shelyag2006} assumed a constant adiabatic exponent [$\Gamma=5/3$] of a perfect gas and then adjusted pressure and density to reach convective stability and hydrostatic equilibrium.
\inlinecite{Parchevsky2007} chose a non-negative profile of $N^2$ and then calculated the corresponding density profile that satisfied hydrostatic equilibrium.
\inlinecite{Schunker2011} constructed Convectively Stable Models (CSM) by taking {Model S} as reference and modifying the sound speed before stabilizing it, such that the mode frequencies of the new stable model are close to those of Model S. 

Stabilization, unfortunately, modifies the solutions for the wave field, and the question arises of how to correct the results that we obtain from the simulations, in order to recover the solutions for the original model of the star. 
We propose here a perturbative approach that numerically corrects 
for the changes in the wave field caused by stabilizing the background model, and approximate the correct solutions of the original unstable model.
This is a step toward direct comparison of synthetic data with data from observations (\eg~observations from the \textit{Helioseismic and Magnetic Imager}: \opencite{Scherrer2012}). 

\section{Proposed Solution: A Perturbative Approach}
\label{sec:generalmethod}

\subsection{Constructing Convectively Stable Background Models}

The linearized equation of motion describing the propagation of acoustic waves inside a star 
has the general form
\begin{equation}
\label{eq:momentumI}
  \mathcal L \vet{\xi}(\vet r,t)  = \vet S (\vet r,t) \,,\quad 
  \mathcal L  =  {\rho} \partial_t^2 +\mathcal H 
\end{equation}
where $\vet r$ is the position vector, $t$ is time, $\mathcal{H}$ is a linear spatial operator associated with the background stellar model, $\vet{\xi}(\vet r,t)$ is the vector wave displacement, and 
$\vet S (\vet r,t)$ is a source function that represents forcing by granulation. 
In the adiabatic case, $\mathcal{H}$ takes the form
\begin{equation}
\label{eq:H0xi}
 \mathcal{H} {\vet \xi} = \nabla p' + {\rho}\left ( {\vet \xi} \cdot {\vet A} -\frac{p'}{\Gamma P}  \right )
                                 {\vet g} - {\rho}{\vet g}' - {\vet F}',
\end{equation}
where primes refer to wave perturbations and the term $\vet F'$ accounts for the interaction of waves with flows and magnetic fields. 
Solutions of Equation (\ref{eq:momentumI}) are uniquely determined once the initial and boundary conditions are set. 
{
 Note that when $\vet F'=0$ the operator $\mathcal H$ is Hermitian and symmetric \cite{1967lynden-bell}. 
}

Let us choose a reference unstable model, \eg~solar Model S,  which is labeled ``ref'' throughout this article.
We construct a convectively stable model defined by the new quantities $\rho_0, P_0,$ and $ \Gamma_0$. 
These quantities are obtained from the original reference model by imposing $N_0^2 \ge 0$.
The simplest choice is to set $N^2_0=0$ where $N^2_\mathrm{ref}$ is negative, but other choices are possible.
We define the differences between the stable and the reference models by
\begin{equation}
\label{eq:stablerhopgamma}
      \Delta\rho  = \rho_\mathrm{ref} -\rho_0 \,,\quad 
        \Delta P  = P_\mathrm{ref} - P_0 \,,\quad 
\Delta\Gamma = \Gamma_\mathrm{ref} - \Gamma_0 \,.
\end{equation}
The difference in the squared buoyancy frequency is then $\Delta N^2= N^2_\mathrm{ref} - N^2_0$.

Stabilization can be achieved in different ways.
In the spherically symmetric case and with the hydrostatic equilibrium condition, stellar models are entirely described by two independent physical quantities (if no flows and no magnetic fields are present): for example, the density [$\rho_0(r)$] and the first adiabatic exponent [$\Gamma_0(r)$]. 
When $\rho_0$ and $\Gamma_0$ are specified, the pressure is given by
\begin{equation}
\label{eq:Hydrostatic}
 \TD{P_0}{r} = - \rho_0(r) g_0(r) ,
\end{equation}
where $g_0(r) >0 $ is the acceleration of gravity, which is fixed by $\rho_0(r)$.

Stabilization by changing $\Gamma_\mathrm{ref}$ is a simple procedure.
On the other hand, changing the density requires solving a nonlinear boundary-value problem, involving Equations (\ref{eq:SchwarzA}) and (\ref{eq:Hydrostatic}) with the new stable $N_0^2$ (\eg~\opencite{Parchevsky2007}).
In the latter case a smart choice of the boundary conditions must be made to preserve the main properties of the star (such as total mass and radius). 
Changing both $\Gamma_\mathrm{ref}$ and $\rho_\mathrm{ref}$ is allowed and desirable, but it is not a straightforward procedure and we do not explore this possibility further in this work.

The linearized equation of motion for the stable model takes the form:
\begin{equation}
\label{eq:momentumH0}
  \mathcal L_0 \vet{\xi}_0(\vet r,t)  = \vet S (\vet r,t) \,,\quad 
  \mathcal L_0  = {\rho_0} \partial_t^2 +\mathcal H_0 \,, 
\end{equation}
where $\mathcal{H}_0$ is the operator associated with the new stable model and $\vet \xi_0$ is the corresponding wave-field solution.

We stress that convective stabilization must be applied consistently with the hypothesis made for the model; 
we also note that density, pressure, and first adiabatic exponent must be changed in Equation (\ref{eq:H0xi}) and all other equations, not only in $\vet A$.

\subsection{First-Order Correction to the Wave Field}
\label{sec:firstorder3d}

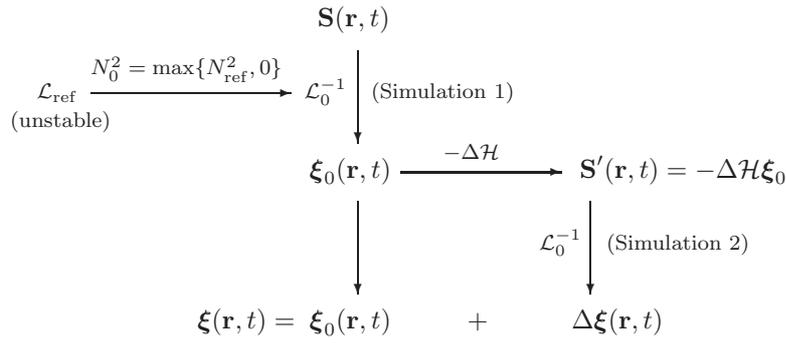
\begin{figure}
\centering
\begin{picture}(130,130)(-50,-10)
\put(-35,114){$\bS(\br,t)$}
\put(-20,105){\vector(0,-1){35}} 
  \put(-15,87){{\scriptsize  (Simulation 1)}}
    \put(-40,87){{\scriptsize $\mathcal L_0^{-1} \;$ }}

\put(-120,96){{\scriptsize $N^2_0=\max\{N^2_\mathrm{ref},0\}$}}
\put(-120,89){\vector(1,0){75}}
\put(-140,87){{\scriptsize $\mathcal L_\mathrm{ref} \;$}}
\put(-150,76){{\scriptsize (unstable)}}
\put(12,64){{\scriptsize $-\Delta\mathcal H$}}
\put(-38,57){$\bxi_0(\br,t)$}    \put(-4,59){\vector(1,0){60}}    \put(63,57){$\vet S'(\br,t)= -\Delta \mathcal H \bxi_0$}
\put(-20,48){\vector(0,-1){35}}                                   \put(67,48){\vector(0,-1){35}}  
                        \put(73,30){{\scriptsize  (Simulation 2)}}
                        \put(48,30){{\scriptsize $\mathcal L_0^{-1} \;$ }}
\put(-38,0){$\bxi_0(\br,t)$}       \put(21,0){$+$} \put(60,0){$\Delta\bxi(\br,t) $}
\put(-80,0){$\vet\xi(\br,t) = \;$}

\end{picture}
 \caption[Sketch of the method]{
    \begin{footnotesize}
     The steps of the proposed method. Here $\mathcal L^{-1}_0$ mathematically represents the operation performed by the simulations. 
     A stable model is built from $\mathcal L_\mathrm{ref}$.
     The solution for the stable model is computed (Simulation 1), and is used to compute the driving source $\vet S'$ (with $-\Delta\mathcal H $).
     A second simulation is run (Simulation 2) to find the correction [$\Delta \bxi $] toward the unstable model.
     Refer to Equations (\ref{eq:momentumH0}) and (\ref{eq:momentumdeltaH}) for the symbols.
    \end{footnotesize}}
 \label{fig:cartoon}
\end{figure}

Assuming that a first simulation to solve Equation (\ref{eq:momentumH0}) is performed and the solution $\vet \xi_0$ for the stable model is computed,
we write the approximate solution [$\vet \xi$] for Equation (\ref{eq:momentumI}) as
\begin{equation}
\label{eq:deltaxi}
 \vet \xi (\vet r,t) = \vet \xi_0 (\vet r,t) + \Delta \vet \xi (\vet r,t),
\end{equation}
where $\Delta {\vet \xi}$ represents the first-order correction to ${\vet \xi}_0$ toward the unstable model. This correction is given by
\begin{equation}
\label{eq:momentumdeltaH}
  \mathcal L_0 \Delta\vet{\xi}(\vet r,t)  = - \Delta\mathcal H \vet\xi_0 (\vet r,t) ,
\end{equation}
 where the operator $\Delta\mathcal H$ is the first correction to the wave operator, obtained by collecting the first-order terms in $\mathcal  L_\mathrm{ref}-\mathcal  L_0$.
In practice, the correction $\Delta \bxi$ is obtained by running a second simulation using the same background model [$\mathcal L_0$] but with a source term $- \Delta\mathcal H \vet\xi_0 (\vet r,t)$. 
Figure \ref{fig:cartoon} sketches the steps of the method. The main advantage of this method is that it is well defined, uses  computational tools, and does not require fine-tuning of the stabilization  
to match the observations (\eg~as in \opencite{Schunker2011}).

{
Applying the correction doubles the computational cost.
  Whether this cost is worth it or not depends on the 
  application. For example, in the future we intend to use the simulations to study the effect 
  of active regions on low-degree modes. Such a small effect (less than 
  a $\mu\mathrm{Hz}$) is at the level of the first-order correction in the background model.
}

\begin{figure}
 \includegraphics[width=\textwidth]{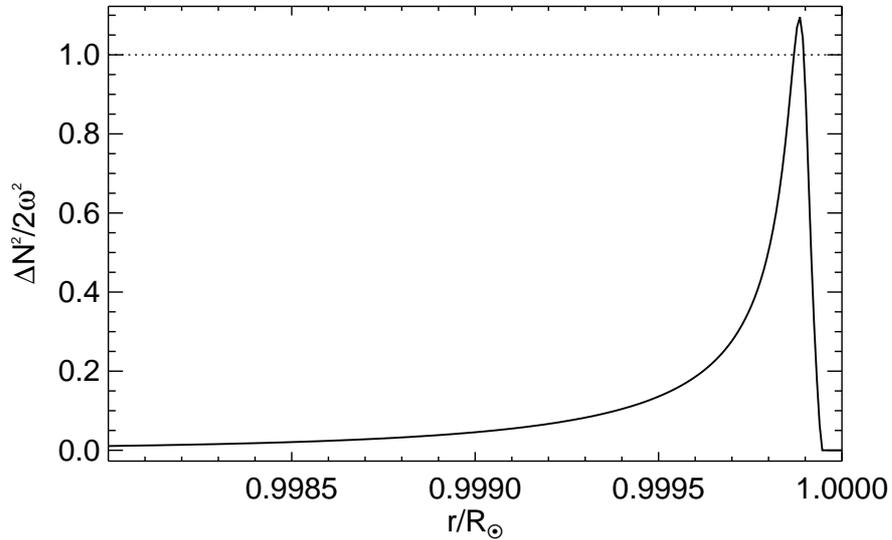}
 \caption[Relative change for $\omega/2\pi=3\,\mathrm{ mHz}$]{
    \begin{footnotesize}
         $ \Delta N^2 / 2 \omega^2 $ in the upper part of the convection zone (solar Model S)
for a frequency of $\omega/2\pi=3 \,\mathrm{ mHz}$.
    \end{footnotesize}}
  \label{fig:dww}
\end{figure}

To assess the validity of the method, one needs to estimate whether the perturbations invoked in Equations~(9) and (10) are weak. To do so, we need to write an approximation for the operator $\Delta \mathcal  H$  as a function of the change in $N^2$. 
By inspection of the wave operator (Equation (5)), we see that an essential term is
\begin{equation}
\Delta\mathcal H \bxi \, \approx \, \rho_0 ( \bxi \cdot \Delta \bA)\vet g = \rho_0\xi_{r} \Delta N^2    \hat{r},
\end{equation}
such that the first-order correction to the mode frequencies may be approximated by (\eg~\opencite{2010Asteroseismo})
\begin{equation}
 \frac{\Delta\omega}{\omega_0}
 \,\approx \,{\int_\odot  
 \Delta N^2   \,\xi_r^2 \rho_0
   \mathrm{d}V \over 
  2 \omega_0^2 \int_\odot \|\bxi\|^2 \rho_0 \mathrm{d}V}  ,
\end{equation}
and the relative correction in the mode frequencies is a weighted average of $ \Delta N^2 / 2\omega^2$. For the first-order perturbation theory to work, we should have $ \Delta N^2 / 2\omega^2 \lesssim 1$. 
Figure \ref{fig:dww} shows $ \Delta N^2 / 2\omega^2 $ for $\omega/2\pi=3\,\mathrm{mHz}$ in the case of solar Model S, which is based on a mixing-length treatment of convection.  This quantity is well below unity throughout the convection zone, except in a localized region near the surface (the highly superadiabatic layer) where it reaches $1.1$ for $\omega/2\pi = 3$~mHz. 
As frequency decreases, $\Delta N^2 / 2\omega^2$ increases; however low-frequency modes are also less sensitive to surface perturbations. Therefore, we expect the first-order perturbation theory to work reasonably well for the full spectrum of solar oscillations. This is shown for particular cases in the following sections.

{
We note that seismic reciprocity \cite{1998Dahlen} is preserved to
first order, since both $\mathcal H_0$ and $\Delta\mathcal H$ are Hermitian and symmetric operators
in the absence of flows and magnetic fields  \cite{1967lynden-bell}.
The concept of seismic reciprocity can be extended to include flows and magnetic fields (see \opencite{Hanasoge2011} and references therein).

Seismic reciprocity is a key property of the adjoint method used to solve the inverse 
  problem in seismology (\eg~\opencite{2005Tromp}; \opencite{Hanasoge2011}).

Modified background models employed by \inlinecite{Hanasoge2006}, \inlinecite{Shelyag2006} and \inlinecite{Parchevsky2007} all satisfy reciprocity.
By contrast, seismic reciprocity is not automatically enforced in 
  the model of \inlinecite{Hartlep2008}, which neglects the term $\bA$ in the momentum equation and in the CSM solar models of \inlinecite{Schunker2011}, which are not hydrostatic.}

\section{Testing the Method in 1D for the Sun}

We tested the method in the 1D hydrodynamic case for the Sun, starting from standard solar Model S \cite{Christensen-Dalsgaard1996}. 
For the test we used the ADIPLS code \cite{Christensen-Dalsgaard2008}, which solves the adiabatic stellar oscillation equations for a spherically symmetric stellar model in hydrostatic equilibrium as an eigenvalue problem (not in the time domain).
This allows one to compute the exact solution for unstable models, and hence directly measure the accuracy of the correction discussed in Section \ref{sec:generalmethod}.

Writing the solution in the form $\vet \xi(\vet r,t)=\vet \xi_{n\ell m} (\vet r) e^{-\rmi\omega_{n\ell} t}$ and setting $\vet S = \vet 0$, we have
\begin{equation}
\label{eq:HXIeigen}
  \mathcal{H} {\vet \xi_{n\ell m}}(\vet r) = \rho\omega^2_{n\ell}{\vet \xi_{n\ell m}}(\vet r) \,,
\end{equation}
where $\omega_{nl}$ is the acoustic mode frequency and $\vet \xi_{n\ell m}(\vet r)$ the corresponding eigenvector displacement (in the following we omit the $n\ell m$ subscripts for clarity).
Each solution is uniquely identified by three integers: the radial order [$n$], the angular degree [$\ell$], and the azimuthal order [$m$], where $|m|\leq \ell$ (in the spherically symmetric case that we consider here the solutions are degenerate in $m$).

For our purpose the operator $\mathcal{H}$ can be written as
\begin{eqnarray}
\label{eq:HXI}
  \mathcal{H}\vet{\xi} =  &-& \nabla (\Gamma P \nabla \cdot {\vet{\xi}}) - \nabla (\vet{\xi} \cdot \nabla P)  \nonumber \\ 
                                 &+& \frac{\nabla P}{\rho} \nabla \cdot{(\rho \vet{\xi})} + \rho \mathrm G \nabla \left ( \int_\odot \frac{\nabla_{\tiny\br'} \cdot {(\rho \vet{\xi})}}{\|\vet{r}-\vet{r}'\|} \mathrm d {V'} \right ) ,
\end{eqnarray}
{where $\rmG$ is the universal gravitational constant;} magnetic fields and flows are not present (see Equation (\ref{eq:H0xi})) and every wave perturbation to pressure and gravity is expressed in terms of $\vet \xi$.

\subsection{Acoustic Modes}

For the test we chose to construct a stable model by only changing $\Gamma$ in Model S to obtain $N^2_0=\max \{N^2_\mathrm{ref},0\} $. 
This was made by setting the first adiabatic exponent [$\Gamma_0$] to
\begin{equation}
 \Gamma_0(r) = \left \{ \begin{array}{llc}
                   & {\displaystyle \Gamma_\mathrm{ref}}(r) \,   & \mathrm{where}\; N^2_\mathrm{ref} \geq 0 \\
                   \\& {\displaystyle \rmd{\,\ln{P_\mathrm{ref}}}} \big{\slash}
                   {\displaystyle  \rmd{\,\ln{\rho_\mathrm{ref}}} } \,  & 
                            \mathrm{where}\; N^2_\mathrm{ref} < 0 \,,
                     \end{array}
            \right .
\end{equation}
where $\rho_\mathrm{ref},\,P_\mathrm{ref},$ and $\Gamma_\mathrm{ref}$ refer to Model S. The density and pressure remained unchanged, \ie~$\rho_\mathrm{ref} =\rho_0$ and  $P_\mathrm{ref} =P_0$.

Solutions for the stable model were computed with ADIPLS, and we calculated the corrections to the eigenfrequencies by using
\begin{equation}
 \label{eq:DeltaHgamma}
 \Delta \mathcal H \vet{\xi}_0= - \nabla (\Delta \Gamma P_0 \nabla \cdot {\vet{\xi}_0})  \,.
\end{equation}
{We note that $\Delta\mathcal H$ is Hermitian and symmetric.}
Given the eigensolutions [$\vet\xi_0,\omega_0^2$] for the stable model, we then calculated the first-order correction to the change in the eigenfrequencies.
\begin{figure}
 \includegraphics[width=\textwidth]{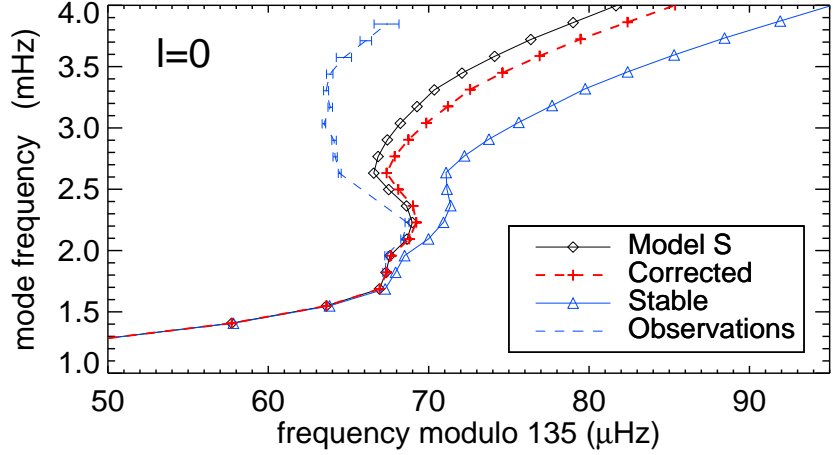}
 \caption[Echelle diagram for radial modes]{
    \begin{footnotesize}
    Echelle diagram showing mode frequencies modulo $135\,\mu\mathrm{Hz}$ for modes with $\ell=0$ and $7\leq n \leq 27$. 
    The first-order correction (red dashed line and crosses) moves the mode frequencies back toward Model S (black solid line and diamonds) from the modified $\Gamma$-stable model (blue solid line and triangles). BiSON data (blue dashed line and error bars) from 108 days of observations starting from 7 February 1997 are plotted for comparison.
    \end{footnotesize}}
  \label{fig:echelle}
\end{figure}

\begin{figure}
 \includegraphics[width=\textwidth]{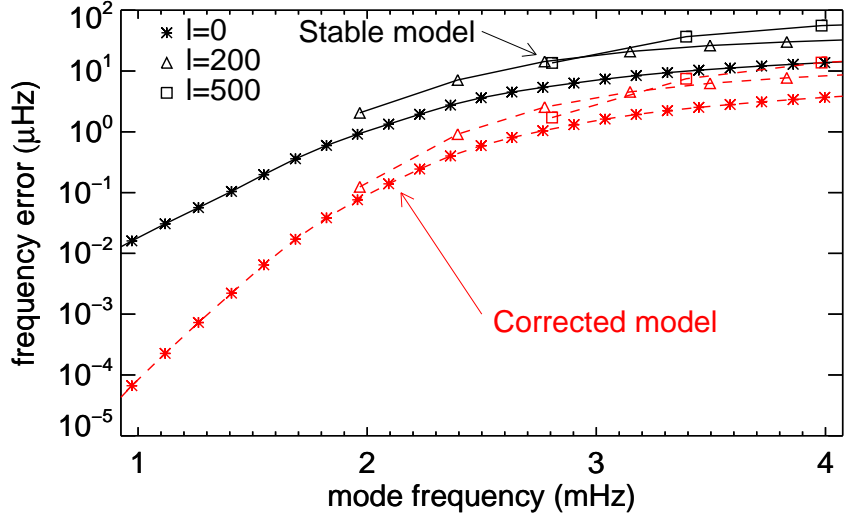}
 \caption[Frequency correction of low-degree modes]{
    \begin{footnotesize}
     Mode-frequency error 
     for {acoustic} modes with {$\ell=0,200,$ and $500$}. Solid black line: difference $(\omega_0-\omega_\mathrm{ref})/2\pi$ between the $\Gamma$-stable model and Model S (reference model). Dashed red line: difference $(\omega_0+\Delta\omega-\omega_\mathrm{ref})/2\pi$ between the corrected frequencies and Model S. 
    \end{footnotesize}}
 \label{fig:correction}
\end{figure}
\begin{figure}
 \includegraphics[width=\textwidth]{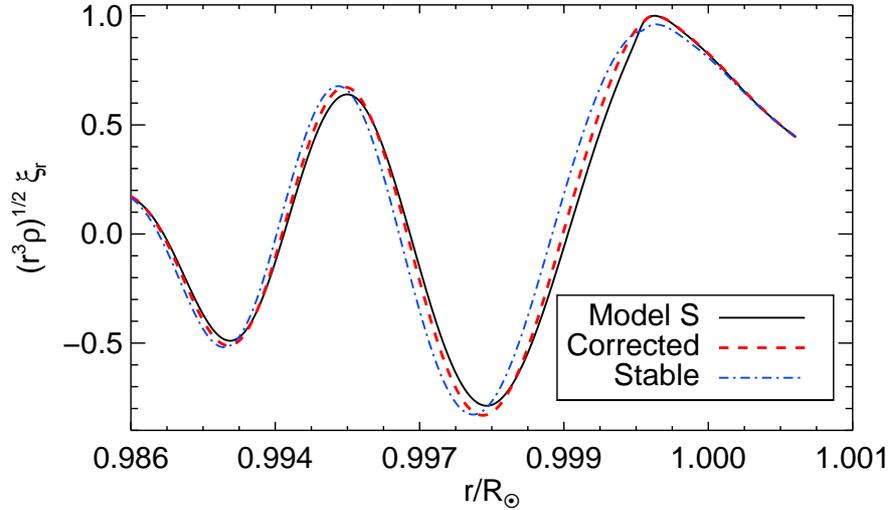}
 \caption[Radial displacement eigenfunction for mode l=500 n=4]{
    \begin{footnotesize}
    {Radial displacement eigenfunction  $(r^3\rho)^{1/2}\xi_r$ for the $\ell=500$ and $n=4$ mode as a function of radius (constant grid spacing in acoustic depth).  
    The solid black line is for Model S, the dash-dotted blue line for the stable model, and the dashed red line for the first-order correction. All three eigenfunctions are normalized with respect to the maximum value of the eigenfunction of Model S.}
    \end{footnotesize}}
\label{fig:eigenxil500n4}
\end{figure}
Test results are shown in Figures \ref{fig:echelle} and \ref{fig:correction}. Figure \ref{fig:echelle} shows the solar \'{e}chelle diagram for the $\ell=0$ modes. The correction moves the mode frequencies from the stable model toward Model S.
Observed frequencies from the \textit{Birmingham Solar\textendash Oscillations Network} (BiSON) \cite{Chaplin2002} are plotted for comparison. 

Figure \ref{fig:correction} shows the plot of mode-frequency differences (for the {$\ell=0,200,$ and $500$} modes) between the stable model and Model S and the residual differences 
between the corrected frequencies and frequencies of Model S.
The correction brings the mode frequencies much closer to the values of the original model:
the difference between the corrected frequencies and those of the reference model is two orders of magnitude smaller than the difference between the stable and the reference model at $1\,\mathrm{ mHz}$. 
The correction is not as efficient as the frequency increases, but still at the level of one order of magnitude at high frequencies.  
That is because as frequency increases acoustic modes are more sensitive to the near surface, where the strongest changes to the model are present.
{
The mode-frequency differences between the stable model and \- Model S increase with $\ell$ since high-degree modes are more sensitive to the surface layers (see $\ell=200$ and $500$ in Figure \ref{fig:correction}).
 The first-order correction reduces these frequency differences by a factor of ten.
In Figure \ref{fig:eigenxil500n4} we display the radial displacement eigenfunctions for the mode $\ell=500$ and $n=4$. 
We see that the first-order correction brings the phase and amplitude of the corrected eigenfunction
closer to those of Model S.}

\subsection{Rotational Sensitivity Kernels}

We furthermore assessed the ability of the method to correct the eigenfunctions by testing with rotational kernels. In the presence of rotation, frequencies are no longer degenerate in the azimuthal order [$m$]. 
 In the case of rotation constant on spheres, the rotational splitting frequency is 
\begin{equation}
 \label{eq:rotsplit}
 S_{n\ell} \equiv \frac{\omega_{n\ell m} - \omega_{n\ell 0}}{m} = \int_0^{R} K(\vet\xi_{n\ell},r)\Omega(r)\mathrm d {r}
\end{equation}
where $\Omega(r)$ is the angular velocity at radius $r \leq R$ and $K$ is the rotational kernel \cite{Hansen1977}. The kernel for mode ($n$, $\ell$) depends on $\bxi_{n\ell}^0$ and the density profile. With ADIPLS we can directly calculate rotational splitting in the case of a rotation profile that only depends on $r$.

The first-order correction  in the rotational splitting frequency as a result of stabilization is
\begin{equation}
 \Delta S_{n\ell}= \int_0^{R} \Delta K_{n\ell}(r)\Omega(r)\mathrm d {r} \,,
\end{equation}
where the perturbation to the kernel can be computed numerically using 
\begin{equation}
 \Delta K_{n\ell}(r) = \lim_{\epsilon \rightarrow 0} \frac{1}{\epsilon}\left [ K(\vet \xi_{n\ell}^\epsilon,r) -K(\vet \xi_{n\ell}^0,r) \right ], 
\end{equation}
where  $\bxi_{n\ell}^\epsilon$ is the eigenvector that solves Equation~(\ref{eq:HXIeigen}) for $\mathcal H_\mathrm{\scriptscriptstyle} = \mathcal H_0 +\epsilon (\mathcal H_\mathrm{ref} -\mathcal H_0)$ and $\epsilon$ is an infinitesimally small parameter. We calculated $\Delta K_{n\ell}$ numerically using $\epsilon=10^{-5}$, in a linear regime where the result is independent of $\epsilon$, within the numerical precision of ADIPLS.

\begin{figure}
 \includegraphics[width=\textwidth]{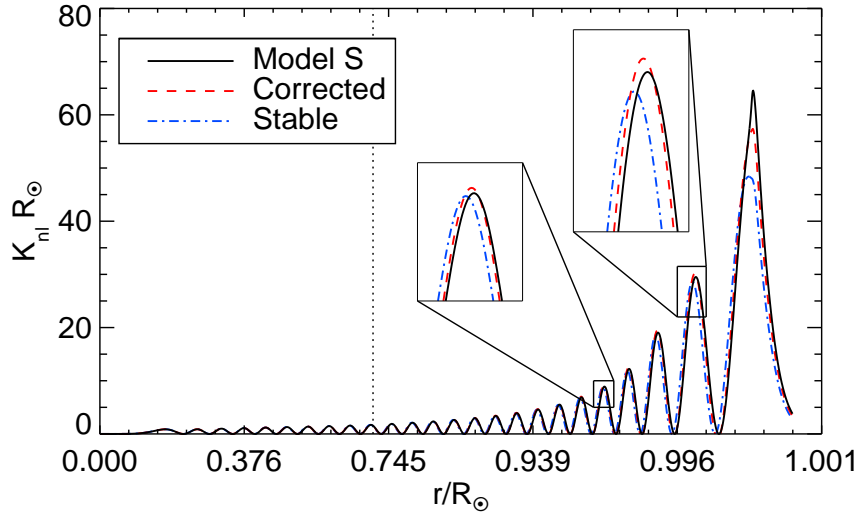}
 \caption[Rotational sensitivity kernel for mode l=1 n=25]{
    \begin{footnotesize}
    Rotational kernels for the $\ell=1$ and $n=25$ mode as a function of radius (constant grid spacing in acoustic depth).  
    The vertical dotted line indicates the location of the base of the convection zone. 
    The solid black line is the kernel for Model S, the dash-dotted blue line for the stable model, and the dashed red line for the first-order correction.
    \end{footnotesize}}
\label{fig:rotkerl1n25}
\end{figure}

Figure \ref{fig:rotkerl1n25} shows the rotational kernel for the $\ell=1,n=25$ mode and the corrected kernel. We see that the phase and amplitude of the corrected kernel are closer to that of the Model S kernel.

To evaluate the accuracy of the correction, we computed the rotational splitting given by Equation (\ref{eq:rotsplit}) in the case of a solid rotation profile of $\Omega/2\pi = 0.5~\mu\mathrm{Hz}$, for the $\ell=1$ modes.
The maximum difference $S_{n\ell}^\mathrm{ref}-(S_{n\ell}^0 + \Delta S_{n\ell})$ between the corrected model and Model S is around $10^{-3} \,\mathrm{nHz}$, while the difference $S_{n\ell}^\mathrm{ref}-S_{n\ell}^0$ between Model S and the stable model is one order of magnitude higher.

\section{Outlook}

We proposed a perturbative approach to run time-domain simulations of wave propagation in a general 3D stellar model. The simulation  was run first using a background model that was convectively stable. First-order perturbation theory was then applied to obtain the corrected wave field.

The method requires that the relative change [$\Delta N$] in the buoyancy frequency between the stable and the unstable model is such that $\Delta N^2 /2 \omega^2 \lesssim 1$, where $\omega$ is the wave angular frequency. Whether this condition is fulfilled depends on the model of convection used. 
In this work we used the solar Model S, which is based on a mixing-length theory (MLT) of convection (by setting $\alpha_P=~1.990$, see \opencite{JCD2008}, Appendix 2 and references therein),
such that $\Delta N^2/2\omega^2<1.1$ at $\omega/2\pi=3$~mHz in the highly superadiabatic layer. Model S does not include the treatment for turbulent pressure.
Other models of convection (including turbulent pressure, MLT with different mixing-length parameters, nonlocal MLT, models from 3D simulations) may result in different superadiabatic gradients (as shown by, \eg~\opencite{Trampedach2010}, Figure 4), leading to either higher or lower values of $\Delta N^2/2\omega^2$. In addition, the peak in the superadiabatic gradient strongly depends on the solar-like star under consideration which, for increasing values of $\log g$, show a decreasing amplitude and an increasing width of the superadiabatic peak (see \opencite{Trampedach2010} Figure 2).

\begin{figure}
 \hspace*{.7 cm}
 \includegraphics[width=.9\textwidth]{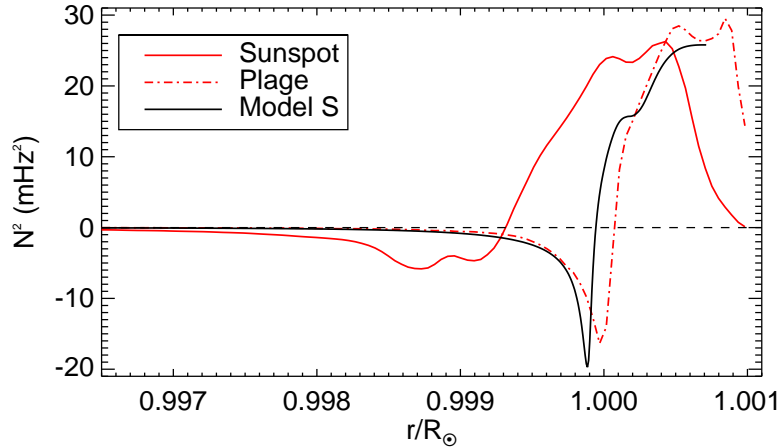}
 \caption[Comparison of buoyancy frequency in sunspot, plage, and Model S]{
    \begin{footnotesize}
    Square of the buoyancy frequency in presence of a sunspot, from the simulations \cite{Braun2012}: in the umbra (red line) (circular average on the first $2$ Mm from the center of the spot), in the plage (red dot dashed line) and from Model S (black line).
    \end{footnotesize}}
    \label{fig:N2SPOT}
\end{figure}

Because we are ultimately interested in running 3D simulations of wave propagation in the presence of magnetic activity, it is of interest to ask about the influence of magnetic fields on superadiabatic gradients in the near surface layers. For that purpose, we measured $N^2$ in the realistic 3D sunspot simulation of \inlinecite{Braun2012}. It follows, using the condition for convective instability in the presence of vertical magnetic fields \cite{Gough1966} [$1/\Gamma -  {\rm d} \ln\rho / {\rm d}\ln P < (1+\gamma_1\beta/2)^{-1}$], that the magnetic field has a stabilizing effect. We thus expect that enforcing $N^2>0$ in the quiet Sun is a sufficient condition for stability in the presence of magnetic activity. Figure~\ref{fig:N2SPOT} shows that the value of $\Delta N^2$ in the sunspot (where the magnetic-field amplitude exceeds $3000$~G) is reduced by a factor of about four. In plage regions (with a magnetic-field amplitude $B \approx100$~G, at a distance of 20~Mm from the center 
of the sunspot),  $\Delta N^2$ is 
only slightly  reduced (Figure~\ref{fig:N2SPOT}).

While more tests are needed, we expect that the proposed approach for performing time-domain simulations of wave propagation will find applications both in local and global helioseismology.

\begin{acks}
The authors acknowledge research funding by the Deutsche  Forsch\-ungsgemeinschaft (DFG) under the grant SFB~963/1 project A18. We used data provided by M. Rempel at the National Center for Atmospheric Research (NCAR). Support for the production of the data was provided by the NASA \textit{Solar Dynamics Observatory} (SDO) Science Center program through grant NNH09AK021 awarded to NCAR and contract NNH09CE41C awarded to NWRA. The National Center for Atmospheric Research is sponsored by the National Science Foundation. 
L.G. acknowledges support from EU FP7 Collaborative Project \textit{Exploitation of Space Data for Innovative Helio- and Asteroseismology} (SPACEINN).
We used data provided by BiSON, funded by the UK Science and Technology Facilities Council (STFC).
We thank Robert Cameron for comments.
\end{acks}

\nocite{*}
\bibliography{bibliography}

\end{document}